\documentclass[aps,prd,preprint,showpacs,superscriptaddress]{revtex4-1}
\usepackage{latexsym}
\usepackage{amsmath,amssymb}
\usepackage{graphicx}
\usepackage{xcolor}
\usepackage{hyperref}     % color hypertext for pdflatex
\hypersetup{colorlinks,%
  citecolor=blue,%
  linkcolor=cyan,%
  pdftex}

\begin{document}

%\preprint{arXiv:yymm.nnnn [hep-th]}

\title{Note on explicit form of entanglement entropy in the RST model}

%%%%%%%%%%%%%%%%%%%% PRD Author Style
\author{Myungseok Eune}%
\email[]{eunems@smu.ac.kr}%
\affiliation{Department of Computer System Engineering, Sangmyung
  University, Cheonan, 330-720, Republic of Korea}%

\author{Yongwan Gim} %
\email[]{yongwan89@sogang.ac.kr} %
\affiliation{Department of Physics, Sogang University, Seoul 121-742,
  Republic of Korea} %

\author{Wontae Kim} %
\email[]{wtkim@sogang.ac.kr} %
\affiliation{Department of Physics, Sogang University, Seoul 121-742,
  Republic of Korea} %
\affiliation{Research Institute for Basic Science, Sogang University,
  Seoul, 121-742, Republic of Korea} %

\date{\today}

\begin{abstract}
  For an evaporating black hole which is a radiation-black hole
  combined system, we express the entanglement entropy and the Page
  time in terms of the conformal time in the RST model.  The entropy
  change of the black hole is nicely written in terms of Hawking flux.
  Integrating the first law of thermodynamics, we can obtain the
  decreasing black hole entropy and the increasing radiation entropy,
  and the entanglement entropy for this system based on the Page
  argument.  We also obtain analytically the critical temperature to
  release black hole information, which corresponds to the Page time,
  and discuss the relation between the conserved total entropy and
  information recovering of the black hole in this model.
\end{abstract}

%\keywords{RST model, Entropy}

\maketitle

%%%%%%%%%%%%%%%%%%%%%%%%%%%%%%%%%%%%%%%%%%%%%%%%

%\section{Introduction}
%\label{sec:intro}

%% Entropies
Bekenstein has suggested that the entropy of a black hole is
proportional to the area of the horizon~\cite{Bekenstein:1972tm,
  Bekenstein:1973ur, Bekenstein:1974ax}, and subsequently Hawking's
discovery has led to the result that the black hole has thermal
radiation with the temperature \(T_{\rm H} = \kappa_{\rm H} / 2\pi
\)~\cite{Hawking:1974sw}, where \(\kappa_{\rm H}\) is the surface
gravity at the event horizon.  It has also been claimed that the black
hole would eventually disappear completely through thermal radiation,
which gives rise to information loss problem~\cite{Hawking:1976ra}.
However, if Hawking radiation plays a role of carrier of information,
information will come out so slowly until the Page
time~\cite{Page:1993wv} when the entanglement entropy becomes maximum
such that the dimension of radiation equals to that of the black hole
in the Hilbert space.  When the dimension of radiation is larger than
that of the black hole, information is naturally contained in
radiation. Moreover, it has been shown that in Ref.~\cite{Page:1993wv}
the above statistical analysis can be realized in the
Callan-Giddings-Harvey-Strominger (CGHS) model~\cite{Callan:1992rs} by
taking into account the classical metric along with the corresponding
constant temperature which is independent of black hole mass so that
radiation does not reflect the back reaction of the geometry.
Actually, the black hole entropy of a two-dimensional black hole with
the back reaction was studied for the static case in
Refs.~\cite{Solodukhin:1994yz, Solodukhin:1995te}.

In this work, we are going to study the entanglement entropy based on
the Page formulation using the RST model~\cite{Russo:1992ax,
  Russo:1992yh} to take into account back reaction of the geometry,
which yields naturally the time-dependent geometry.  The essential
difficulty is to identify the time-dependent temperature which is
quite awkward in standard thermodynamics. So we would like to assume
that a radiation-black hole combined system is in equilibrium at each
time such that the radiation temperature measured by the fixed
observer at the future null infinity is identified with the black hole
temperature.  Then, the thermodynamic first law is also read off from
the differential form of the energy conservation
law~\cite{Kim:1995wr}, so that the entropy change of the black hole is
neatly written in terms of Hawking flux.  Integrating the first law of
thermodynamics, we can obtain the decreasing black hole entropy and
the increasing radiation entropy, and the entanglement entropy based
on the Page argument~\cite{Page:1993wv}. So, the total entropy is
always constant while the total information is not conserved locally
because of the time-dependent entanglement entropy; however, it is
expected that the total information is recovered after complete
evaporation of the black hole.
%\section{Black hole entropy in the RST model}
%\label{sec:rst}
%% 2D dilaton gravity

Now, let us start with the RST model given by the
action~\cite{Russo:1992ax}
\begin{align}
  I &= I_{\rm DG} + I_f + I_{\rm PL} + I_{\rm
    corr}, \label{action:total}
\end{align}
with
\begin{align}
  I_{\rm DG} &= \frac{1}{2\pi} \int d^2 x \sqrt{-g}\, e^{-2\phi}
  \left[R + 4(\nabla\phi)^2 + 4\lambda^2
  \right], \label{action:DG} \\
  I_f &= \frac{1}{2\pi} \int d^2 x \sqrt{-g} \left[- \frac12
    \sum_{i=1}^{N} (\nabla f_i)^2 \right], \label{action:f} \\
  I_{\rm PL} &= \frac{\kappa}{2\pi} \int d^2 x \sqrt{-g}
  \left[-\frac14 R \frac{1}{\Box} R\right], \label{action:PL} \\
  I_{\rm corr} &= \frac{\kappa}{2\pi} \int d^2 x \sqrt{-g}
  \left[-\frac12 \phi R \right], \label{action:corr}
\end{align}
where \(\kappa = (N-24)/12\) which can be positive by taking into
account the ghost decoupling term~\cite{Strominger:1992th} and
$\lambda$ is a cosmological constant. Eq.~\eqref{action:corr} is added
to obtain an exact black hole solution and it is reduced to the CGHS
model without this term~\cite{Callan:1992rs}.  From the
action~\eqref{action:total}, the equations of motion are given by
\(\Box f_i =0\), \(e^{-2\phi} \left[R - 4 (\nabla\phi)^2 + 4\Box\phi +
  4\lambda^2\right] + \frac{\kappa}{4} R =0\), and \( G_{\mu\nu} =
T_{\mu\nu}^f + T_{\mu\nu}^{\rm qt} \), where \( G_{\mu\nu} \equiv
\frac{2\pi}{\sqrt{-g}} \frac{\delta}{\delta g^{\mu\nu}} \left(I_{\rm
    DG} + I_{\rm corr}\right) = 2e^{-2\phi} \left[\nabla_\mu
  \nabla_\nu \phi + g_{\mu\nu} \left((\nabla \phi)^2 - \Box\phi -
    \lambda^2\right) \right] + \frac{\kappa}{2} \left(\nabla_\mu
  \nabla_\nu \phi - g_{\mu\nu} \Box\phi \right)\). The energy-momentum
tensors for matter are defined as
\begin{align}
  T_{\mu\nu}^f &\equiv - \frac{2\pi}{\sqrt{-g}} \frac{\delta
    I_f}{\delta g^{\mu\nu}} \notag \\
  &= \sum_{i=1}^N \left[\frac12 \nabla_\mu f_i \nabla_\nu f_i -
    \frac14 g_{\mu\nu} (\nabla f_i)^2 \right], \label{T:cov:cl} \\
  T_{\mu\nu}^{\rm qt} &\equiv - \frac{2\pi}{\sqrt{-g}} \frac{\delta
    I_{\rm PL}}{\delta g^{\mu\nu}}. \label{T:cov:qt}
\end{align}
It can be checked that the dilaton improved Bianchi identity for the
RST model is satisfied, \textit{i.e.,} \(\nabla^\mu G_{\mu\nu} = 0\),
which yields covariant conservation relations for classical matter and
quantum matter as \(\nabla^\mu T_{\mu\nu}^f = 0\) = \(\nabla^\mu
T_{\mu\nu}^{\rm qt} = 0\).  Thus, the definitions of the classical and
the quantum energy-momentum tensors given by Eqs.~\eqref{T:cov:cl} and
\eqref{T:cov:qt} in the RST model are compatible with those of the
CGHS model as long as covariant conservation relations are concerned.
In the conformal gauge given by \(ds^2 = e^{2\rho} dx^+ dx^-\),
% Eqs.~\eqref{eom:phi:cov} and \eqref{eom:f:cov} are written as
% \begin{align}
%   &e^{-2\phi} \left(\partial_+ \partial_- \rho + 2 \partial_+
%     \phi \partial_- \phi - 2 \partial_+ \partial_- \phi + \frac12
%     \lambda^2 e^{2\rho} \right) +
%   \frac{\kappa}{4} \partial_+ \partial_- \rho =
%   0, \label{eom:phi:conf} \\
%   &\partial_+ \partial_- f_i = 0, \label{eom:f:conf}
% \end{align}
% and Eq.~\eqref{G:cov} becomes
% \begin{align}
%   G_{\pm\pm} &= 2e^{-2\phi} \left(\partial_\pm^2 \phi - 2 \partial_\pm
%     \rho \partial_\pm \phi \right), \label{G:++} \\
%   G_{\pm\mp} &= e^{-2\phi} \left(4\partial_+\phi \partial_- \phi -
%     2\partial_+ \partial_- \phi + \lambda^2 e^{2\rho}
%   \right). \label{G:+-}
% \end{align}
the classical energy-momentum tensor~\eqref{T:cov:cl} is written as \(
T_{\pm\pm}^f = \frac12 \sum_{i=1}^N (\partial_\pm f_i)^2\) and \(
T_{\pm\mp}^f =0\) and the quantum energy-momentum
tensor~\eqref{T:cov:qt} is given by \(T_{\pm\pm}^{\rm qt} = \kappa
\left[\partial_\pm^2 \rho - (\partial_\pm \rho)^2 - t_\pm (x^\pm)
\right]\) and \(T_{\pm\mp}^{\rm qt} = -\kappa \partial_+ \partial_-
\rho\), which agrees with the quantum energy-momentum tensor
introduced in the CGHS model~\cite{Callan:1992rs}. The unknown
functions \(t_\pm(x^\pm)\) reflect the nonlocal property of the
effective action. Of course, one may define the energy-momentum tensor
for matter in a different way by including the contribution from Eq.
\eqref{action:corr} because it is actually not unique.  However, it
can be well-defined at asymptotic future null infinity since it can be
expressed by only the boundary function as $T_{\pm\pm}^{\rm qt} =
-\kappa t_{\pm} $ when we consider Hawking radiation in that region.

 By introducing new variables given by \(\chi =
\sqrt\kappa \rho - (\sqrt\kappa/2) \phi + e^{-2\phi}/\sqrt\kappa \)
and \(\Omega= (\sqrt\kappa/2) \phi + e^{-2\phi}/\sqrt\kappa\) for
simplicity, the action~\eqref{action:total} can be written
as~\cite{Russo:1992ax, Russo:1992yh},
\begin{align}
  I = \frac{1}{\pi} \int d^2 x \left[- \partial_+
    \chi \partial_- \chi + \partial_+
    \Omega \partial_- \Omega + \lambda^2 e^{(2/\sqrt\kappa)(\chi - \Omega)}
    + \frac12 \sum_{i=1}^N \partial_+ f_i \partial_- f_i
  \right], \label{S:+-}
\end{align}
and the two constraints are given by \(\kappa t_\pm = -
\partial_\pm \chi \partial_\pm \chi + \sqrt\kappa \partial_\pm^2 \chi +
\partial_\pm \Omega \partial_\pm \Omega + \frac12
\sum_{i=1}^N \partial_\pm f_i \partial_\pm f_i\).  The equations of
motion derived from the action~\eqref{S:+-} can be exactly solved. In
the Kruskal coordinates where \(\chi = \Omega\), the evaporating black
hole formed by an incoming shock wave of~ \(T^f_{++} = [M/(\lambda
x^+_0)] \delta(x^+ - x^+_0)\) is described by the solution of $\Omega
(x^+, x^-)= - \lambda^2 x^+ x^- / \sqrt\kappa - \frac{\sqrt\kappa}{4}
\ln(-\lambda^2 x^+ x^-) - \frac{M}{\lambda \sqrt\kappa x^+_0} (x^+ -
x^+_0) \Theta (x^+ - x^+_0)$, where the linear dilaton vacuum is
chosen for \(x^+< x^+_0\). An asymptotically static coordinate can be
obtained from the coordinate transformations defined by \(x^+ =
(1/\lambda) e^{\lambda \sigma^+} \) and \(x^- = - (1/\lambda)
e^{-\lambda \sigma^-} - ( M/\lambda^2) e^{-\lambda \sigma^+_0}
\Theta(\sigma^+ - \sigma^+_0)\), where \(\sigma^+_0 = \lambda^{-1}
\ln(\lambda x^+_0) \).

\begin{figure}[htb]
  \centering
  \includegraphics[width=0.48\textwidth]{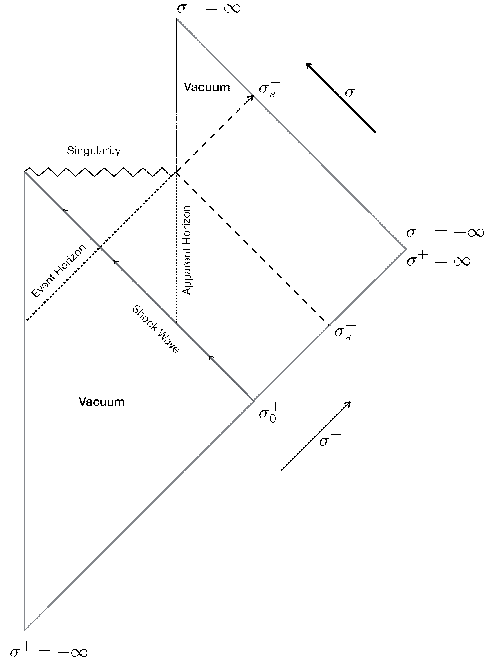}
  \caption{It shows the Penrose diagram of the black hole formed by a
    shock wave at $\sigma^+ = \sigma^+_0$.}
  \label{fig:penrose}
\end{figure}

Note that the RST model is known to be quantum-mechanically
inconsistent after appearance of the naked singularity
~\cite{Strominger:1994tn}.  The curvature singularity and apparent
horizon collide in a finite proper time and the singularity is naked
after the two have merged~\cite{Russo:1992ax}. In order to avoid the
naked singularity, a vacuum state can be patched at the intersection
point ($\sigma^+_s, \sigma^-_s$) of the singularity curve and the
apparent horizon as shown in the Penrose diagram of
Fig.~\ref{fig:penrose}, where the intersection point is given by
\(\sigma^-_s = \sigma^+_0 + \lambda^{-1} \ln \left[(\lambda/M)
  \left(\exp(4M/(\kappa\lambda)) - 1 \right) \right]\) and
\(\sigma^+_s = \sigma^-_s + \lambda^{-1} \ln (\kappa/4)\)
\cite{Russo:1992ax,Russo:1992yh,Kim:1995wr,Strominger:1994tn}. However,
this patching procedure requires the thunderpop energy which is the
negative classical energy emanated from the black hole. So we are going to mainly
discuss the entanglement entropy of the RST model before the negative
Bondi mass appears.

%%%%%%%%%%%%%

On the other hand, from the covariant conservation law, one can get
the ordinary conserved quantity by expanding the metric and dilaton
fields around the linear dilaton vacuum. Then, the linearized equation
of motion becomes \(G^{(1)}_{ \mu \nu} = T^f_{\mu \nu} - G^{(2)}_{\mu
  \nu}\)~\cite{Weinberg:1972book}, where $T^f_{\mu \nu}$ is a
classical energy-momentum tensor, $G^{(1)}_{\mu \nu}$ is the linear
perturbed part of $G_{\mu \nu}$, and $G^{(2)}_{\mu \nu}$ is the rest.
Then, one can choose the time and space coordinate so that it is easy
to show that the linearized equation of motion identically satisfies
the ordinary conservation law, $\partial_\mu G^{(1) \mu 0} = 0,$ by
the use of the linearized Bianchi identity~\cite{Kim:1995jta}.  It
implies that the current defined as $J^\mu = T_f^{\mu 0} - G^{(2) \mu
  0}$ satisfies the ordinary conservation law $\partial_\mu J^{\mu}=
0$.  Thus we can define the Bondi mass $B(\sigma^-)$ which is the
energy evaluated along the null line \cite{Bondi:1962px}, $B(\sigma^-)
= (1/2)\int_{-\infty}^\infty d\sigma^+ G^{(1) - 0} (\sigma^+, \sigma^-
)$ while the ADM mass is calculated at the spatial infinity as $E_{\rm
  ADM}(t) = \int_{-\infty}^\infty dq G^{(1)00} (t,q)$
\cite{Arnowitt:1962hi}.  Using the integrated form of the linearized
equation of motion, after some calculations the difference between the
ADM mass and the Bondi mass can be obtained as~\cite{Kim:1995wr}
\begin{equation}
  \label{bt}
  E_{\rm ADM}(t)-B(\sigma^-) = \int_{-\infty}^{\sigma^-} d\sigma^-
  \left. \left( T_{--}^f +T_{--}^{\rm qt} \right) \right|_{\sigma^+ \to
    \infty}.
\end{equation}
Note that the classical infalling energy-momentum tensor does not
exist since it cannot appear in the asymptotic future null infinity.
However, in the RST model there may exist the classical out-going
negative energy density called the thunderpop energy at
$\sigma^{-}_s$.  From now on, we will consider radiation-hole combined
system before the thunderpop appears, and then the conformal time is
restricted to $-\infty < \sigma <\sigma^{-}_s$.  It means that we can
take vanishing out-going classical energy-momentum tensor in this
analysis.

Next, the integrated Hawking flux is given by $ H(\sigma^-) =
\int_{-\infty}^{\sigma^-} d\sigma^- h(\sigma^-)$, where the Hawking
flux is $h(\sigma^-) = \left. T_{--}^{\rm qt} \right|_{\sigma^+ \to
  \infty}$.  The Hawking flux is simply reduced to the boundary
function as $h(\sigma^-) = - t_- (\sigma^-)$ since \(\sigma^\pm\) is a
quasi-static coordinate system at infinity, and so the fields approach
the linear dilaton vacuum at \(\sigma^+ \to \infty\).  In this black
hole, the Hawking radiation is written as $ h(\sigma^-) =
(\kappa\lambda^2/4) [1 - (1 + (M/\lambda) e^{\lambda (\sigma^- -
  \sigma^+_0)} )^{-2} ]$~\cite{Callan:1992rs}.  Note that the Bondi
mass is the remaining energy after quantum-mechanical Hawking
radiation has been emitted from the system.  So it is plausible to
regard the Bondi mass as a black hole mass in the quantum back-reacted
model.  From Eq. \eqref{bt}, we can get the conservation law
as~\cite{Kim:1995wr}
\begin{align}
  B(\sigma^-) + H(\sigma^-)=M, \label{E:BH}
\end{align}
where $M$ is the ADM mass.  The energy can be conserved in this
evaporating black hole system so that the Bondi energy plus the
Hawking radiation should be equal to the initial infalling energy by
the scalar fields.

Now, we will assume the radiation-black hole combined system as a
thermal equilibrium system for each conformal time $\sigma^-$ in order
to apply the thermodynamic first law. Let us first relate the Hawking
flux to the Hawking temperature in analogy with the static
case~\cite{KeskiVakkuri:1993xi}, then one can read off the black
hole temperature \(T(\sigma^-)\) from the Hawking radiation by
identifying
\begin{equation}
  h(\sigma^-) = \kappa \pi^2 T^2(\sigma^-),
\end{equation}
which yields
\begin{align}
  T(\sigma) &= \frac{\lambda}{2\pi} \left[1 - \frac{1}{\left(1 +
        \frac{M}{\lambda} e^{\lambda (\sigma^- - \sigma^+_0)}
      \right)^2} \right]^{1/2}. \label{T:def}
\end{align}
Note that it vanishes at $\sigma^- \to -\infty$ since the black hole
did not radiate yet and the well-known Hawking temperature is
recovered as \(T_{\rm H} = \lambda/2\pi\) at \(\sigma^- \to \infty\)
which is compatible with the previous static
results~\cite{Page:1993wv,KeskiVakkuri:1993xi}.  Using the
differential form of the energy conservation law~\eqref{E:BH}, the
change of the black hole entropy can be written as
\begin{align}
  \Delta S_h & = S_h(\sigma^-) - S_h^0 = \int \frac{dB}{T} \notag \\
  &= - \pi \sqrt{\kappa} \int_{-\infty}^{\sigma^-} d\sigma^-
  \sqrt{h(\sigma^-)}, \label{entropy:change}
\end{align}
where \(S_h^0\) denotes the entropy of the black hole at \(\sigma^-
\to -\infty\).  The entropy change is essentially due to Hawking
radiation such that the entropy of the black hole is decreasing.  From
Eq. \eqref{entropy:change} the entropy is calculated as
\begin{align}
  S_h (\sigma^-) &=\frac{2\pi M}{\lambda} - \frac{\pi\kappa}{2}
  \left[\sec^{-1} \gamma(\sigma^-) + \ln\left(\gamma(\sigma^-) +
      \sqrt{\gamma^2(\sigma^-) -1}\right) \right], \label{S:bh}
\end{align}
where \(\gamma(\sigma^-) = 1 + (M/\lambda) e^{\lambda(\sigma^- -
  \sigma^+_0)} \) and we employed the fact that the entropy of the
black hole is given by \(S_h^0 = 2\pi M/\lambda\) at the initial time
of \(\sigma^- \to -\infty\) since the entropy of the black hole starts
with the maximum thermal entropy of the area law, and at the same time
the Hawking temperature~\eqref{T:def} is zero.  As time goes on, the
black hole entropy is decreasing according to the increasing Hawking
temperature which amounts to \(T_{\rm H} = \lambda/2\pi\) at
\(\sigma^- \to \infty\). Note that in the conventional thermodynamic
analysis, the black hole entropy and the temperature are given as $S =
2\pi M/\lambda$, and $T_{\rm H} = \lambda/2\pi $.

%\section{Discussion}
%\label{sec:discuss}
On the other hand, for a system consisting of the black hole subsystem
and the radiation subsystem, the entanglement entropy for $S_h, S_r \gg 1$ is given by the
Page argument as~\cite{Page:1993wv}
\begin{align}
   S_{\rm ent} &\simeq \left\{
    \begin{array}{ll}
      %\displaystyle
      S_r - \frac12 e^{S_r-S_h} &\qquad
      \mathrm{for}\ S_r\le S_h \\
      \vspace{2mm}
      %\displaystyle
      S_h - \frac12 e^{S_h-S_r} & \qquad
      \mathrm{for}\ S_r\ge S_h
    \end{array}
    \right., \label{S:ent}
\end{align}
where \(S_h \) and \(S_r\) are the black hole
entropy and the radiation entropy, respectively.  Note that the total
entropy of the system is preserved such that it is given as \(S_r +
S_h = 2\pi M/\lambda\). The entanglement entropy~\eqref{S:ent} has a
maximum value at the Page time when the black hole emits a half of its
initial Bekenstein-Hawking entropy, $i.e.$, \(S_r =\pi M/\lambda\).
Using Eq.~\eqref{S:bh}, we can write the entanglement entropy
explicitly in terms of \(\sigma^-\), and it becomes
\begin{align}
  S_{\rm ent} (\sigma^-)&\simeq \frac{\pi\kappa}{2} \Bigg[\sec^{-1}
  \gamma(\sigma^-) + \ln\left(\gamma(\sigma^-) +
    \sqrt{\gamma^2(\sigma^-) -1}\right) \Bigg] \notag \\
  &\quad - \frac12 \left[ \left(\gamma(\sigma^-) +
      \sqrt{\gamma^2(\sigma^-) -1}\right) \exp \left(\sec^{-1}
      \gamma(\sigma^-) - \frac{2 M}{\kappa \lambda}\right)
  \right]^{\pi \kappa}
\end{align}
for \(\sigma^- \le \sigma^-_c\) and
\begin{align}
  S_{\rm ent}  (\sigma^-)&\simeq \frac{2\pi M}{\lambda} - \frac{\pi\kappa}{2}
  \left[\sec^{-1}
    \gamma(\sigma^-) + \ln\left(\gamma(\sigma^-) +
      \sqrt{\gamma^2(\sigma^-) -1}\right) \right] \notag \\
  &\quad -  \frac12 \left[ \left(\gamma(\sigma^-) +
      \sqrt{\gamma^2(\sigma^-) -1}\right) \exp \left(\sec^{-1}
      \gamma(\sigma^-) - \frac{2 M}{\kappa \lambda}\right)
  \right]^{-\pi \kappa}
\end{align}
for \(\sigma^- \ge \sigma^-_c\).  Note that the entanglement entropy
becomes maximum at the conformal time of $\sigma^{-}_c$ which comes
from the maximization of the entanglement entropy formally given in
the closed form of \(\gamma(\sigma^-_c) \cos [\ln(\gamma(\sigma^-_c) +
\sqrt{\gamma^2(\sigma^-_c) - 1}) - 2M/(\kappa\lambda) ] =1\).
\begin{figure}[htb]
  \centering
  \includegraphics[width=0.5\textwidth]{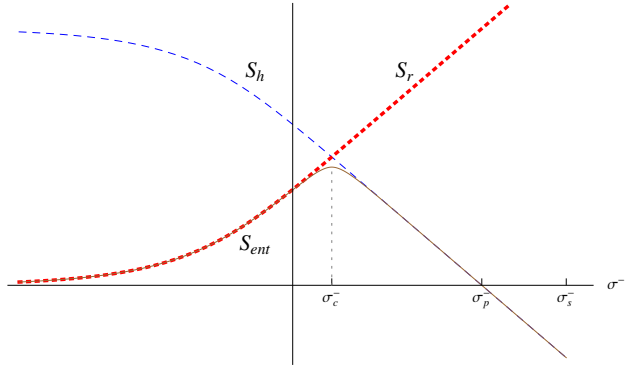}
  \caption{The solid, the dashed, and the thick-dotted lines show the
    behaviors of the entanglement entropy \(S_{\rm {ent}}\), the black
    hole entropy \(S_h\), and the thermodynamic radiation entropy
    \(S_r\), respectively.  The entanglement entropy has a maximum at
    $\sigma^-_c$ and it vanishes at $\sigma^-_p$.}
  \label{fig:I:S.ent:S.r::sigma}
\end{figure}
That point is just the Page time expressed by the conformal time since
the radiation entropy is the same with the black hole entropy as shown
in Fig.~\ref{fig:I:S.ent:S.r::sigma}. The radiation entropy is
monotonically increasing while the black hole entropy is monotonically
decreasing, and their sum is constant.

By the way, there is a deficiency in this calculation that the black
hole entropy is negative for $\sigma^- > \sigma^-_p$ because the Bondi
mass in the RST model is negative due to the surplus Hawking radiation
after $\sigma^{-}_p$ \cite{Strominger:1994tn} so that the present
calculations are meaningful before $\sigma^-_p$.  Moreover, the
expression for the entanglement entropy based on the Page argument is
valid only for many degrees of freedom as was noticed below
\eqref{S:ent} such as $S_h, S_r \gg 1$.  So, it seems to be
inappropriate to discuss beyond the end point of the entropy in our
formulation, and the calculation of the entanglement entropy becomes a
good approximation around $\sigma^-_c$.  One more thing to be
mentioned is that we distinguished the definition of the entropy
depending on the subsystem: the black hole entropy is defined by
employing the Bondi mass, which is plausible in that the entropy
change of the black hole should be negative because the black hole
radiates while the entropy change of radiation should be positive
because Hawking radiation is increased monotonically.

As for the naked singularity of the black hole, the black hole can generally form
 a singularity.
  However, as seen from the original work \cite{Page:1993wv}, the black hole system was assumed to have many degrees of freedom
   such as $m, n \gg 1$ in order to formulate the system and derive the explicit form of
   the average value of the entanglement entropy of Eq. (\ref{S:ent}). It means that even in spite of
   the small black hole, it should have many degrees of freedom in this formulation so that
    the black hole does not lose its mass completely and then the naked singularity is no
    more concerned. Based on this argument, we have employed the same entropy formula
    in the present RST model so that our result is also valid only for the many degrees of
     freedom just at the conformal time $\sigma^- \ll \sigma_p^-$ for which $S_h \gg 1$.
     Therefore, the entanglement entropy turns out to be well-defined only around $\sigma_c^-$ in Fig. \ref{fig:I:S.ent:S.r::sigma} except the extreme limits of very small degrees of freedom. Furthermore, the
     advantage of the RST model in Ref. \cite{Russo:1992ax} is that it has been designed to be free from the naked
     singularity because the flat metric can be patched with the black hole metric when
     the singularity forms at $\sigma_s^-$ as seen from Fig. \ref{fig:penrose}.

As a result, we have obtained the decreasing black hole entropy, the
increasing radiation entropy, the entanglement entropy, and the Page
time in terms of the conformal time in the exactly soluble RST model.
Moreover, we can find a Page temperature at the Page time since
$\sigma^{-}_c$ was identified so that it becomes formally
$T(\sigma^-_c)$ from Eq. \eqref{T:def}.  In other words, information
is significantly leased above the critical temperature of
$T({\sigma^{-}_c})$.

 In Refs. \cite{Myers:1994sg,  Hayward:1994dw}, the black hole entropy and increase theorem related
to the second law of black hole thermodynamics have been studied for
the RST model, and we would like to mention some differences between
our work and them. First, the system in our work was divided into two
subsystems so that the black hole has the black hole entropy and
radiation has the radiation entropy, respectively, while there appears
only a single system and the single entropy to define the black hole
system in the previous works.  Additionally, the entanglement entropy
in this work has been defined throughout correlation between the two
subsystems, so that the entropy in Refs. \cite{Myers:1994sg,
  Hayward:1994dw} behaves like not the entanglement entropy but the
radiation entropy in our work in the sense that it is always
increasing as time goes on and the entropy change is always positive,
which guarantees the second law of black hole thermodynamics.

In the original work done by Page in Ref. \cite{Page:1993wv},
the system was divided into two subsystems; one is for the black hole with dimension $m$
and the other is for radiation with dimension $n$. The most important assumption is that
these subsystems form a total system in a pure state in a Hilbert space of fixed dimension $mn$.
 It means that the total entropy is constant, and consequently $\Delta S=0$. Following this assumption in Ref. \cite{Page:1993wv},
  we also assumed that the total entropy should be constant in order to realize the Page argument in the RST model. In particular, for the case of $\Delta S>0$,
   the information  will be eventually lost like ordinary thermal systems. Therefore, the
   requirement of the fixed total entropy is a sort of constraint based on the hypothesis
   that no information is lost in black hole formation and evaporation as was claimed in Ref. \cite{Page:1993wv}.

 %%%%%% Acknowledgments %%%%%%%%%
\begin{acknowledgments}
 W. Kim was supported by the National Research Foundation of Korea(NRF) grant funded by the Korea government (MOE) (2010-0008359).
\end{acknowledgments}

%%%%%%%%%%%%%%%%%%%%%%%%%%%%%%%%%%%%%%%%%%%%%%%%
%%%%%%%%%%%%%%%             References         %%%%%%%%%%%%%%%%
%%%%%%%%%%%%%%%%%%%%%%%%%%%%%%%%%%%%%%%%%%%%%%%%

%\bibliographystyle{mybib}
\bibliographystyle{apsrev4-1} % PRD
\bibliography{references}

\begin{thebibliography}{21}%
\makeatletter
\providecommand \@ifxundefined [1]{%
 \@ifx{#1\undefined}
}%
\providecommand \@ifnum [1]{%
 \ifnum #1\expandafter \@firstoftwo
 \else \expandafter \@secondoftwo
 \fi
}%
\providecommand \@ifx [1]{%
 \ifx #1\expandafter \@firstoftwo
 \else \expandafter \@secondoftwo
 \fi
}%
\providecommand \natexlab [1]{#1}%
\providecommand \enquote  [1]{``#1''}%
\providecommand \bibnamefont  [1]{#1}%
\providecommand \bibfnamefont [1]{#1}%
\providecommand \citenamefont [1]{#1}%
\providecommand \href@noop [0]{\@secondoftwo}%
\providecommand \href [0]{\begingroup \@sanitize@url \@href}%
\providecommand \@href[1]{\@@startlink{#1}\@@href}%
\providecommand \@@href[1]{\endgroup#1\@@endlink}%
\providecommand \@sanitize@url [0]{\catcode `\\12\catcode `\$12\catcode
  `\&12\catcode `\#12\catcode `\^12\catcode `\_12\catcode `\%12\relax}%
\providecommand \@@startlink[1]{}%
\providecommand \@@endlink[0]{}%
\providecommand \url  [0]{\begingroup\@sanitize@url \@url }%
\providecommand \@url [1]{\endgroup\@href {#1}{\urlprefix }}%
\providecommand \urlprefix  [0]{URL }%
\providecommand \Eprint [0]{\href }%
\providecommand \doibase [0]{http://dx.doi.org/}%
\providecommand \selectlanguage [0]{\@gobble}%
\providecommand \bibinfo  [0]{\@secondoftwo}%
\providecommand \bibfield  [0]{\@secondoftwo}%
\providecommand \translation [1]{[#1]}%
\providecommand \BibitemOpen [0]{}%
\providecommand \bibitemStop [0]{}%
\providecommand \bibitemNoStop [0]{.\EOS\space}%
\providecommand \EOS [0]{\spacefactor3000\relax}%
\providecommand \BibitemShut  [1]{\csname bibitem#1\endcsname}%
\let\auto@bib@innerbib\@empty
%</preamble>
\bibitem [{\citenamefont {Bekenstein}(1972)}]{Bekenstein:1972tm}%
  \BibitemOpen
  \bibfield  {author} {\bibinfo {author} {\bibfnamefont {J.~D.}\ \bibnamefont
  {Bekenstein}},\ }\href {\doibase 10.1007/BF02757029} {\bibfield  {journal}
  {\bibinfo  {journal} {Lett. Nuovo Cim.}\ }\textbf {\bibinfo {volume} {4}},\
  \bibinfo {pages} {737} (\bibinfo {year} {1972})}\BibitemShut {NoStop}%
%%CITATION = NCLTA,4,737;%%
\bibitem [{\citenamefont {Bekenstein}(1973)}]{Bekenstein:1973ur}%
  \BibitemOpen
  \bibfield  {author} {\bibinfo {author} {\bibfnamefont {J.~D.}\ \bibnamefont
  {Bekenstein}},\ }\href {\doibase 10.1103/PhysRevD.7.2333} {\bibfield
  {journal} {\bibinfo  {journal} {Phys. Rev. D}\ }\textbf {\bibinfo {volume}
  {7}},\ \bibinfo {pages} {2333} (\bibinfo {year} {1973})}\BibitemShut
  {NoStop}%
%%CITATION = PHRVA,D7,2333;%%
\bibitem [{\citenamefont {Bekenstein}(1974)}]{Bekenstein:1974ax}%
  \BibitemOpen
  \bibfield  {author} {\bibinfo {author} {\bibfnamefont {J.~D.}\ \bibnamefont
  {Bekenstein}},\ }\href {\doibase 10.1103/PhysRevD.9.3292} {\bibfield
  {journal} {\bibinfo  {journal} {Phys. Rev. D}\ }\textbf {\bibinfo {volume}
  {9}},\ \bibinfo {pages} {3292} (\bibinfo {year} {1974})}\BibitemShut
  {NoStop}%
%%CITATION = PHRVA,D9,3292;%%
\bibitem [{\citenamefont {Hawking}(1975)}]{Hawking:1974sw}%
  \BibitemOpen
  \bibfield  {author} {\bibinfo {author} {\bibfnamefont {S.}~\bibnamefont
  {Hawking}},\ }\href {\doibase 10.1007/BF02345020} {\bibfield  {journal}
  {\bibinfo  {journal} {Commun. Math. Phys.}\ }\textbf {\bibinfo {volume}
  {43}},\ \bibinfo {pages} {199} (\bibinfo {year} {1975})}\BibitemShut
  {NoStop}%
%%CITATION = CMPHA,43,199;%%
\bibitem [{\citenamefont {Hawking}(1976)}]{Hawking:1976ra}%
  \BibitemOpen
  \bibfield  {author} {\bibinfo {author} {\bibfnamefont {S.}~\bibnamefont
  {Hawking}},\ }\href {\doibase 10.1103/PhysRevD.14.2460} {\bibfield  {journal}
  {\bibinfo  {journal} {Phys. Rev. D}\ }\textbf {\bibinfo {volume} {14}},\
  \bibinfo {pages} {2460} (\bibinfo {year} {1976})}\BibitemShut {NoStop}%
%%CITATION = PHRVA,D14,2460;%%
\bibitem [{\citenamefont {Page}(1993)}]{Page:1993wv}%
  \BibitemOpen
  \bibfield  {author} {\bibinfo {author} {\bibfnamefont {D.~N.}\ \bibnamefont
  {Page}},\ }\href {\doibase 10.1103/PhysRevLett.71.3743} {\bibfield  {journal}
  {\bibinfo  {journal} {Phys. Rev. Lett.}\ }\textbf {\bibinfo {volume} {71}},\
  \bibinfo {pages} {3743} (\bibinfo {year} {1993})},\ \Eprint
  {http://arxiv.org/abs/hep-th/9306083} {hep-th/9306083} \BibitemShut {NoStop}%
%%CITATION = HEP-TH/9306083;%%
\bibitem [{\citenamefont {Callan}\ \emph {et~al.}(1992)\citenamefont {Callan},
  \citenamefont {Giddings}, \citenamefont {Harvey},\ and\ \citenamefont
  {Strominger}}]{Callan:1992rs}%
  \BibitemOpen
  \bibfield  {author} {\bibinfo {author} {\bibfnamefont {C.~G.}\ \bibnamefont
  {Callan}}, \bibinfo {author} {\bibfnamefont {S.~B.}\ \bibnamefont
  {Giddings}}, \bibinfo {author} {\bibfnamefont {J.~A.}\ \bibnamefont
  {Harvey}}, \ and\ \bibinfo {author} {\bibfnamefont {A.}~\bibnamefont
  {Strominger}},\ }\href {\doibase 10.1103/PhysRevD.45.R1005} {\bibfield
  {journal} {\bibinfo  {journal} {Phys. Rev. D}\ }\textbf {\bibinfo {volume}
  {45}},\ \bibinfo {pages} {1005} (\bibinfo {year} {1992})},\ \Eprint
  {http://arxiv.org/abs/hep-th/9111056} {hep-th/9111056} \BibitemShut {NoStop}%
%%CITATION = HEP-TH/9111056;%%
\bibitem [{\citenamefont {Solodukhin}(1995)}]{Solodukhin:1994yz}%
  \BibitemOpen
  \bibfield  {author} {\bibinfo {author} {\bibfnamefont {S.~N.}\ \bibnamefont
  {Solodukhin}},\ }\href {\doibase 10.1103/PhysRevD.51.609} {\bibfield
  {journal} {\bibinfo  {journal} {Phys. Rev. D}\ }\textbf {\bibinfo {volume}
  {51}},\ \bibinfo {pages} {609} (\bibinfo {year} {1995})},\ \Eprint
  {http://arxiv.org/abs/hep-th/9407001} {hep-th/9407001} \BibitemShut {NoStop}%
%%CITATION = HEP-TH/9407001;%%
\bibitem [{\citenamefont {Solodukhin}(1996)}]{Solodukhin:1995te}%
  \BibitemOpen
  \bibfield  {author} {\bibinfo {author} {\bibfnamefont {S.~N.}\ \bibnamefont
  {Solodukhin}},\ }\href {\doibase 10.1103/PhysRevD.53.824} {\bibfield
  {journal} {\bibinfo  {journal} {Phys. Rev. D}\ }\textbf {\bibinfo {volume}
  {53}},\ \bibinfo {pages} {824} (\bibinfo {year} {1996})},\ \Eprint
  {http://arxiv.org/abs/hep-th/9506206} {hep-th/9506206} \BibitemShut {NoStop}%
%%CITATION = HEP-TH/9506206;%%
\bibitem [{\citenamefont {Russo}\ \emph {et~al.}(1992)\citenamefont {Russo},
  \citenamefont {Susskind},\ and\ \citenamefont {Thorlacius}}]{Russo:1992ax}%
  \BibitemOpen
  \bibfield  {author} {\bibinfo {author} {\bibfnamefont {J.~G.}\ \bibnamefont
  {Russo}}, \bibinfo {author} {\bibfnamefont {L.}~\bibnamefont {Susskind}}, \
  and\ \bibinfo {author} {\bibfnamefont {L.}~\bibnamefont {Thorlacius}},\
  }\href {\doibase 10.1103/PhysRevD.46.3444} {\bibfield  {journal} {\bibinfo
  {journal} {Phys. Rev. D}\ }\textbf {\bibinfo {volume} {46}},\ \bibinfo
  {pages} {3444} (\bibinfo {year} {1992})},\ \Eprint
  {http://arxiv.org/abs/hep-th/9206070} {hep-th/9206070} \BibitemShut {NoStop}%
%%CITATION = HEP-TH/9206070;%%
\bibitem [{\citenamefont {Russo}\ \emph {et~al.}(1993)\citenamefont {Russo},
  \citenamefont {Susskind},\ and\ \citenamefont {Thorlacius}}]{Russo:1992yh}%
  \BibitemOpen
  \bibfield  {author} {\bibinfo {author} {\bibfnamefont {J.~G.}\ \bibnamefont
  {Russo}}, \bibinfo {author} {\bibfnamefont {L.}~\bibnamefont {Susskind}}, \
  and\ \bibinfo {author} {\bibfnamefont {L.}~\bibnamefont {Thorlacius}},\
  }\href {\doibase 10.1103/PhysRevD.47.533} {\bibfield  {journal} {\bibinfo
  {journal} {Phys. Rev. D}\ }\textbf {\bibinfo {volume} {47}},\ \bibinfo
  {pages} {533} (\bibinfo {year} {1993})},\ \Eprint
  {http://arxiv.org/abs/hep-th/9209012} {hep-th/9209012} \BibitemShut {NoStop}%
%%CITATION = HEP-TH/9209012;%%
\bibitem [{\citenamefont {Kim}\ and\ \citenamefont {Lee}(1995)}]{Kim:1995wr}%
  \BibitemOpen
  \bibfield  {author} {\bibinfo {author} {\bibfnamefont {W.~T.}\ \bibnamefont
  {Kim}}\ and\ \bibinfo {author} {\bibfnamefont {J.}~\bibnamefont {Lee}},\
  }\href {\doibase 10.1103/PhysRevD.52.2232} {\bibfield  {journal} {\bibinfo
  {journal} {Phys. Rev. D}\ }\textbf {\bibinfo {volume} {52}},\ \bibinfo
  {pages} {2232} (\bibinfo {year} {1995})},\ \Eprint
  {http://arxiv.org/abs/hep-th/9502115} {hep-th/9502115} \BibitemShut {NoStop}%
%%CITATION = HEP-TH/9502115;%%
\bibitem [{\citenamefont {Strominger}(1992)}]{Strominger:1992th}%
  \BibitemOpen
  \bibfield  {author} {\bibinfo {author} {\bibfnamefont {A.}~\bibnamefont
  {Strominger}},\ }\href {\doibase 10.1103/PhysRevD.46.4396} {\bibfield
  {journal} {\bibinfo  {journal} {Phys. Rev. D}\ }\textbf {\bibinfo {volume}
  {46}},\ \bibinfo {pages} {4396} (\bibinfo {year} {1992})},\ \Eprint
  {http://arxiv.org/abs/hep-th/9205028} {hep-th/9205028} \BibitemShut {NoStop}%
%%CITATION = HEP-TH/9205028;%%
\bibitem [{\citenamefont {Strominger}(1994)}]{Strominger:1994tn}%
  \BibitemOpen
  \bibfield  {author} {\bibinfo {author} {\bibfnamefont {A.}~\bibnamefont
  {Strominger}},\ }\href@noop {} {\enquote {\bibinfo {title} {{Les Houches
  lectures on black holes}},}\ } (\bibinfo {year} {1994}),\ \bibinfo {note}
  {talk given at Conference: C94-08-02.1},\ \Eprint
  {http://arxiv.org/abs/hep-th/9501071} {hep-th/9501071} \BibitemShut {NoStop}%
%%CITATION = HEP-TH/9501071;%%
\bibitem [{\citenamefont {Weinberg}(1972)}]{Weinberg:1972book}%
  \BibitemOpen
  \bibfield  {author} {\bibinfo {author} {\bibfnamefont {S.}~\bibnamefont
  {Weinberg}},\ }\href@noop {} {\emph {\bibinfo {title} {{Gravitation and
  Cosmology: Principles and Applications of the General Theory of
  Relativity}}}}\ (\bibinfo  {publisher} {Wiley},\ \bibinfo {address} {New
  York},\ \bibinfo {year} {1972})\BibitemShut {NoStop}%
\bibitem [{\citenamefont {Kim}\ and\ \citenamefont {Lee}(1996)}]{Kim:1995jta}%
  \BibitemOpen
  \bibfield  {author} {\bibinfo {author} {\bibfnamefont {W.~T.}\ \bibnamefont
  {Kim}}\ and\ \bibinfo {author} {\bibfnamefont {J.}~\bibnamefont {Lee}},\
  }\href {\doibase 10.1142/S0217751X96000250} {\bibfield  {journal} {\bibinfo
  {journal} {Int. J. Mod. Phys. A}\ }\textbf {\bibinfo {volume} {11}},\
  \bibinfo {pages} {553} (\bibinfo {year} {1996})},\ \Eprint
  {http://arxiv.org/abs/hep-th/9502078} {hep-th/9502078} \BibitemShut {NoStop}%
%%CITATION = HEP-TH/9502078;%%
\bibitem [{\citenamefont {Bondi}\ \emph {et~al.}(1962)\citenamefont {Bondi},
  \citenamefont {van~der Burg},\ and\ \citenamefont {Metzner}}]{Bondi:1962px}%
  \BibitemOpen
  \bibfield  {author} {\bibinfo {author} {\bibfnamefont {H.}~\bibnamefont
  {Bondi}}, \bibinfo {author} {\bibfnamefont {M.}~\bibnamefont {van~der Burg}},
  \ and\ \bibinfo {author} {\bibfnamefont {A.}~\bibnamefont {Metzner}},\ }\href
  {\doibase 10.1098/rspa.1962.0161} {\bibfield  {journal} {\bibinfo  {journal}
  {Proc. Roy. Soc. Lond. A}\ }\textbf {\bibinfo {volume} {269}},\ \bibinfo
  {pages} {21} (\bibinfo {year} {1962})}\BibitemShut {NoStop}%
%%CITATION = PRSLA,A269,21;%%
\bibitem [{\citenamefont {Arnowitt}\ \emph {et~al.}(2008)\citenamefont
  {Arnowitt}, \citenamefont {Deser},\ and\ \citenamefont
  {Misner}}]{Arnowitt:1962hi}%
  \BibitemOpen
  \bibfield  {author} {\bibinfo {author} {\bibfnamefont {R.~L.}\ \bibnamefont
  {Arnowitt}}, \bibinfo {author} {\bibfnamefont {S.}~\bibnamefont {Deser}}, \
  and\ \bibinfo {author} {\bibfnamefont {C.~W.}\ \bibnamefont {Misner}},\
  }\href {\doibase 10.1007/s10714-008-0661-1} {\bibfield  {journal} {\bibinfo
  {journal} {Gen. Rel. Grav.}\ }\textbf {\bibinfo {volume} {40}},\ \bibinfo
  {pages} {1997} (\bibinfo {year} {2008})},\ \Eprint
  {http://arxiv.org/abs/gr-qc/0405109} {gr-qc/0405109} \BibitemShut {NoStop}%
%%CITATION = GR-QC/0405109;%%
\bibitem [{\citenamefont {Keski-Vakkuri}\ and\ \citenamefont
  {Mathur}(1994)}]{KeskiVakkuri:1993xi}%
  \BibitemOpen
  \bibfield  {author} {\bibinfo {author} {\bibfnamefont {E.}~\bibnamefont
  {Keski-Vakkuri}}\ and\ \bibinfo {author} {\bibfnamefont {S.~D.}\ \bibnamefont
  {Mathur}},\ }\href {\doibase 10.1103/PhysRevD.50.917} {\bibfield  {journal}
  {\bibinfo  {journal} {Phys. Rev. D}\ }\textbf {\bibinfo {volume} {50}},\
  \bibinfo {pages} {917} (\bibinfo {year} {1994})},\ \Eprint
  {http://arxiv.org/abs/hep-th/9312194} {hep-th/9312194} \BibitemShut {NoStop}%
%%CITATION = HEP-TH/9312194;%%
\bibitem [{\citenamefont {Myers}(1994)}]{Myers:1994sg}%
  \BibitemOpen
  \bibfield  {author} {\bibinfo {author} {\bibfnamefont {R.~C.}\ \bibnamefont
  {Myers}},\ }\href {\doibase 10.1103/PhysRevD.50.6412} {\bibfield  {journal}
  {\bibinfo  {journal} {Phys.Rev.D}\ }\textbf {\bibinfo {volume} {50}},\
  \bibinfo {pages} {6412} (\bibinfo {year} {1994})},\ \Eprint
  {http://arxiv.org/abs/hep-th/9405162} {hep-th/9405162} \BibitemShut {NoStop}%
%%CITATION = HEP-TH/9405162;%%
\bibitem [{\citenamefont {Hayward}(1995)}]{Hayward:1994dw}%
  \BibitemOpen
  \bibfield  {author} {\bibinfo {author} {\bibfnamefont {J.~D.}\ \bibnamefont
  {Hayward}},\ }\href {\doibase 10.1103/PhysRevD.52.2239} {\bibfield  {journal}
  {\bibinfo  {journal} {Phys.Rev.D}\ }\textbf {\bibinfo {volume} {52}},\
  \bibinfo {pages} {2239} (\bibinfo {year} {1995})},\ \Eprint
  {http://arxiv.org/abs/gr-qc/9412065} {gr-qc/9412065} \BibitemShut {NoStop}%
%%CITATION = GR-QC/9412065;%%
\end{thebibliography}%

%%%%%%%%%%%%%%%%%%%%%%%%%%%%%%%%%%%%%%%%%%%%%%%%%%%
%%%%%%%%%%%%%%%             References (.bib)      %%%%%%%%%%%%%%%%
%%%%%%%%%%%%%%%%%%%%%%%%%%%%%%%%%%%%%%%%%%%%%%%%%%%

%\begin{filecontents}{references.bib}
%
%
%\end{filecontents}

\end{document}